# Research Novelty in Information Systems Journals After ChatGPT: Differences Across Institutional Language Contexts

Ali Safari[1] Sahar Babaei[2]

[1]Department of Information Technology and Decision Sciences, University of North Texas, Denton, Texas, USA

[2]Texas Tech University, Lubbock, Texas, USA

## Abstract

Large language models are increasingly used in scholarly work, yet it remains unclear whether their productivity gains are accompanied by changes in research novelty. We examine how relative abstract-level semantic novelty in Information Systems journals changed after ChatGPT became widely available and whether this change differed across institutional language contexts. We analyze 13,847 articles published from 2020 to 2025 in 44 A* and A Information Systems journals. Using SPECTER2 representations of titles and abstracts, we measure each article's semantic distance from its nearest recent predecessors and estimate a comparative pre/post model. Articles whose first authors were affiliated with institutions in non-English-dominant countries show a 0.176 standard deviation larger post-2022 decline in relative semantic novelty than articles from English-dominant affiliations, equivalent to about 7 percentile points. The pattern is similar across several alternative specifications, although the balanced-author estimate is less precise. We interpret this finding through a tension in generative AI-supported knowledge work. GenAI can widen access to prior knowledge and support new combinations, but it can also make established frames easier to reproduce. Because individual LLM use is not observed, the result identifies a heterogeneous post-2022 shift rather than an effect of LLM adoption. The study extends research on LLMs and scholarly productivity by shifting attention from publication counts to the semantic positioning of published articles and by showing that post-2022 change differs across institutional contexts.

## 1 Introduction

Large language models have changed how knowledge workers search, write, summarize, and solve problems. Experimental and field evidence shows that these tools can reduce task time and improve output quality, especially for less experienced workers and people who begin with lower performance (Brynjolfsson et al. 2025; Noy and Zhang 2023). Academic work is also changing. Kwon and Yang (2025), using more than 218,000 computer science papers, report an increase in publication output after the release of ChatGPT. Their study raises an important follow-up question: does more output also mean more scientific advancement?

This question matters because research value cannot be judged only by the number of papers produced. Scientific progress often comes from connecting existing ideas in uncommon ways (Uzzi et al. 2013). Yet novelty is difficult and risky. Novel papers may receive weaker recognition in the short term even when they later become influential (Wang et al. 2017). Research systems also tend to become more concentrated around established work as fields grow (Chu and Evans 2021), and scientific outputs have become less disruptive over long periods (Park et al. 2023). If LLMs make established language and familiar research frames easier to use, higher productivity could occur alongside greater similarity in published work.

At the same time, there is no simple reason to expect generative AI to reduce novelty. GenAI can help researchers search large bodies of knowledge, compare explanations, and identify connections that would otherwise be difficult to see. From a knowledge creation perspective, it can support the combination of explicit knowledge across many sources and act as an active tool in the creation process (Alavi et al. 2024). Research on AI use before the LLM period also finds a positive association between AI-related research activity and scientific novelty, especially when AI supports knowledge recombination (Liu et al. 2025). These studies suggest a knowledge-expansion path in which AI helps researchers reach beyond their existing information limits.

The same technology can also produce a standardization path. GenAI learns from accumulated knowledge and can carry old patterns into new outputs. Benbya et al. (2024) explain that this dependence on prior knowledge can support discovery, but it can also repeat outdated practices and make new content echo the limits of its inputs. Experimental work shows a related pattern: access to AI-generated ideas can improve individual creative output while making the full set of outputs more similar (Doshi and Hauser 2024). In academic work, Gopal et al. (2025) also warn that over-delegation can create stylistic homogenization and weaken the distinct voice and judgment of researchers. The effect of generative AI on novelty is therefore not built into the technology. It depends on how the tool enters the knowledge process.

We study this tension in published Information Systems research. The empirical question is whether relative semantic novelty changed after ChatGPT became widely available, and whether this change differed across institutional language contexts. We focus on articles whose first authors are affiliated with institutions in English-dominant countries and those affiliated with institutions in non-English-dominant countries. This comparison is relevant because LLM-based writing support may have different value where scholars face different institutional language conditions. However, affiliation country is not a measure of a researcher's first language, English ability, or LLM use. It is a broad feature of the institutional setting and must be interpreted that way.

We use SPECTER2 embeddings to represent the title and abstract of each article and calculate its distance from the nearest articles published in the prior two years. We call this measure abstract-level semantic novelty. A higher value means that the article's title and abstract occupy a more distant position from recent IS articles in the embedding space. It does not measure all parts of scientific novelty. In particular, it cannot directly capture methodological innovation, theoretical depth, data originality, or novelty that appears only in the full text.

The analysis covers 13,847 articles published from 2020 through 2025 in 44 A* and A IS journals. The main comparative pre/post estimate is a Post x Non-EDA coefficient of -0.176 ($p < 0.001$). In the post-2022 period, the relative semantic novelty of articles in the non-English-dominant affiliation group declined by about 0.18 standard deviations compared with the English-dominant affiliation group. The pattern is stable in several alternative specifications. The estimate from a balanced panel of repeat authors is similar in direction but is not statistically precise. Because all researchers could access LLMs after their release, and because we do not observe actual LLM use, the design does not establish that LLM adoption caused the shift.

This paper makes three connected contributions. It extends research on LLMs and academic productivity by examining the semantic position of published articles rather than publication counts alone. It also separates two forms of AI in science that are often mixed together. AI used as a research method may expand what researchers can analyze and discover, while LLMs used for search, framing, and writing may also make common representations easier to reproduce. Finally, the paper offers a measured account of heterogeneity across institutional language contexts. The observed pattern is important, but it does not show that one group uses LLMs more, writes weaker papers, or has less research ability. It shows that the groups followed different post-2022 paths on one bounded measure of semantic novelty.

# 2 Theoretical Background

## 2.1 Scientific novelty and semantic position

Scientific novelty usually involves a departure from recent or established knowledge. One important view treats novelty as uncommon recombination. High-impact research often connects familiar knowledge with combinations that are unusual in the existing literature (Uzzi et al. 2013). A related text-based view asks whether a paper is semantically distant from prior work. Embedding methods make it possible to represent a scientific document as a point in a high-dimensional space and compare that position with earlier documents (Cohan et al. 2020; Shibayama et al. 2021; Singh et al. 2023).

These views overlap, but they are not identical. Citation-based recombination captures how a paper brings prior sources together. Semantic distance captures how the paper presents its topic and contribution in text. Disruption measures whether later work uses the focal paper instead of its predecessors. Each measure represents a different part of novelty. The current study focuses only on semantic distance in titles and abstracts. We therefore use the more specific term “abstract-level semantic novelty” and avoid treating the measure as a complete test of scientific innovation.

This narrower definition is especially important in the LLM setting. Changes in title and abstract language could reflect changes in research ideas, changes in how authors frame those ideas, or both. SPECTER2 is trained for scientific document representation and uses citation-informed learning, which makes it more suitable than a general sentence model for mapping research documents (Singh et al. 2023). Still, the embedding cannot fully separate idea content from the language used to express that content. The outcome should be understood as the semantic position communicated through the title and abstract.

### 2.2 GenAI and two paths in knowledge work

Research on GenAI in knowledge work points to two competing paths. The first is knowledge expansion. GenAI can process many sources, retrieve information quickly, and help users combine explicit knowledge. Alavi et al. (2024) describe GenAI as an active tool in knowledge combination because it can aggregate, summarize, and identify patterns across large amounts of information. Benbya et al. (2024) also note that GenAI can uncover relationships that may not be clear to human decision makers and can stimulate new directions. For researchers, this can widen the set of ideas considered during literature search, problem framing, and analysis.

The second path is standardization. The model produces output from patterns learned from accumulated data. This can make common concepts, citations, structures, and forms of explanation easy to recover. Benbya et al. (2024) describe the risk that past knowledge artifacts are carried into new content, which can repeat outdated practices or create circular arguments. Alavi et al. (2024) similarly note that an overabundance of AI-provided information may weaken a user's own knowledge seeking and creative capacity. In research practice, this problem may become stronger when users accept the first clear and plausible response instead of comparing several explanations or returning to original sources.

The two paths can operate at the same time. A researcher may reach more information but explore a smaller set of frames. A paper may become clearer while moving closer to common language in the field. An individual output may improve while the distribution of outputs becomes less diverse. Doshi and Hauser (2024) show this last pattern in a creative writing experiment: AI access improved evaluated creativity at the individual level but increased similarity across stories. The tension is therefore not simply productivity versus quality. It is broader access to knowledge versus repeated use of the same accessible patterns.

This distinction also helps explain why earlier findings about AI and scientific innovation do not directly answer the current question. Liu et al. (2025) study more than 32 million papers through 2019 and find that papers with greater use of AI terms or AI knowledge are associated with higher novelty and disruption. In that setting, AI is often part of the research method or knowledge base. Their data end before the wide availability of ChatGPT. The present study examines a later period in which a general-purpose language model may enter many parts of research production, including writing, summarizing, and framing. AI as a method for discovery and LLM support in the production of a paper are related but different forms of use.

### 2.3 Institutional language context as a source of heterogeneity

LLMs can reduce the time and effort needed to write and revise academic English. This benefit may be especially meaningful in institutional settings where English is not the main public and professional language. Yet language context is difficult to measure. Affiliation country does not identify a person's first language, education, writing skill, or actual use of a language model. A scholar working in the United States may have learned English later in life, while a native English speaker may work at an institution in a country classified as non-English-dominant.

Evidence on LLM-assisted scholarly writing also shows why this distinction requires care. Kobak et al. (2025) find large differences in detectable LLM-related vocabulary across countries, fields, and journals, but they explain that these differences may reflect editing behavior and publication timing as well as actual use. Native English speakers may be better able to remove unusual model language, which makes some forms of use harder to detect. Institutional language context should therefore be treated as a source of possible heterogeneity, not as an adoption measure.

Kwon and Yang (2025) use a language-group comparison to study post-ChatGPT publication output in computer science. They find that publication growth did not close the output gap between native and non-native English-speaking scholars. Their study also calls for article-level analysis of whether the additional output represents scientific advancement. We extend that question from publication volume to abstract-level semantic novelty and use affiliation context rather than claiming individual language status.

The standardization path gives a directional expectation. If LLM-based language and framing support changes the research production process more in non-English-dominant institutional contexts, the post-2022 semantic shift may also be larger in that group. The current design cannot test the use mechanism, but it can test whether the group trajectories differ.

**H1:** The post-2022 change in relative abstract-level semantic novelty is more negative for articles whose first authors are affiliated with institutions in non-English-dominant countries than for articles whose first authors are affiliated with institutions in English-dominant countries.

# 3 Method

## 3.1 Data and sample

The sample includes articles published from 2020 through 2025 in 44 Information Systems journals rated A* or A by the Australian Business Deans Council. The list includes the AIS Senior Scholars' Basket of Eight and 36 additional journals, including Decision Support Systems, Information & Management, and Computers in Human Behavior. Bibliographic records were retrieved from Web of Science. After requiring a non-missing abstract, the analytic sample contains 13,847 articles.

We identify the first author's institutional affiliation at the time of publication. Articles are classified as English-dominant affiliation (EDA) when the first author's institution is in the United States, United Kingdom, Canada, Australia, New Zealand, or Ireland. All other affiliation countries are classified as non-English-dominant affiliation (non-EDA). The EDA group contains 8,414 articles and the non-EDA group contains 5,433 articles. This variable describes the institutional language context. It does not identify individual language proficiency, nationality, or LLM use.

## 3.2 Measuring abstract-level semantic novelty

We measure semantic novelty in three steps. We first combine each article's title and abstract and represent the document with a 768-dimensional SPECTER2 embedding (Singh et al. 2023). SPECTER2 is designed for scientific documents and builds on citation-informed document representation (Cohan et al. 2020).

We then compare each article published in year $t$ with articles published in years $t$ - 2 and $t$ - 1. The main measure is based on cosine distance from the focal article to its 10 nearest prior neighbors. A larger distance means that the focal title and abstract are less similar to the closest recent articles in the IS sample. We use alternative values of $k$ equal to 5, 15, and 20 in the robustness analysis.

Finally, we standardize the novelty measure within each publication year. The number of prior articles available for comparison grows over time, which can make nearest-neighbor distances decline mechanically as the embedding space becomes denser. Within-year standardization removes this mechanical difference and keeps the article's relative position in that year's novelty distribution. This choice has an important consequence: the outcome cannot be used to estimate an aggregate rise or decline in novelty across years. It can only show whether groups changed relative to each other within the yearly distribution.

### 3.3 Comparative pre/post specification

We estimate a comparative pre/post model around the public release of ChatGPT in late 2022. The Post indicator equals 1 for articles published in 2023 or later. The main model is:

$$Novelty^{z}_{it} = \beta_0 + \beta_1 Post_t + \beta_2 NonEDA_i + \beta_3 (Post_t \times NonEDA_i) + X_{it}\delta + \varepsilon_{it}$$

The outcome is within-year standardized abstract-level semantic novelty. The coefficient of interest, beta 3, captures whether the post-2022 change differs between the non-EDA and EDA affiliation groups. The reported controls include the log number of coauthors, a seniority indicator, a Tier 1 journal indicator, a COVID-period indicator, and journal fixed effects across model specifications. The reported standard errors are heteroskedasticity-consistent HC1 standard errors.

This specification is similar to a difference-in-differences model, but its interpretation is more limited than a standard treatment-control design. ChatGPT became available to both groups. We do not observe individual adoption, intensity of use, or the stage of research in which an LLM was used. The interaction therefore estimates a heterogeneous post-2022 shift, not a treatment effect of LLM adoption.

Publication timing creates another limit. An article published in 2023 may have been designed, written, or accepted before ChatGPT became available. A 2024 break is used as a robustness check, but publication year remains an imperfect measure of exposure timing. The direction of the resulting bias cannot be known because publication lag, changes in author composition, journal decisions, and topic changes may operate together.

We examine pre-period group differences with year-specific interactions using 2022 as the reference year. The reported interaction estimates for 2020 and 2021 are close to zero. We also use a false 2022 break within the 2020 to 2022 pre-period. These tests provide limited support for similar pre-period paths, but they cannot establish the parallel-trends assumption because only two earlier years are available.

### 3.4 Robustness analyses

The reported robustness analyses change the number of nearest neighbors, move the break to 2024, remove the 2020 and 2021 COVID years, and restrict the sample to 1,079 authors who publish in both the pre- and post-2022 periods. The balanced panel reduces changes in author composition but also reduces the number of articles from 13,847 to 3,239.

## 4 Results

### 4.1 Descriptive statistics

Table 1 presents the reported descriptive statistics. The mean raw cosine distance to the 10 nearest prior neighbors is 0.060, with a standard deviation of 0.015. The within-year standardized outcome has a mean of zero and a standard deviation of one by construction. The sample is 39 percent non-EDA, 36 percent senior first authors, and 13 percent Tier 1 journal articles.

**Table 1. Descriptive statistics**

| Variable | N | Mean | SD | Min | Max |
|---|---|---|---|---|---|
| Raw semantic novelty, k = 10 | 13,847 | 0.060 | 0.015 | 0.015 | 0.189 |
| Within-year standardized novelty | 13,847 | 0.000 | 1.000 | -3.12 | 8.84 |
| Post, 2023 or later | 13,847 | 0.52 | 0.50 | 0 | 1 |
| Non-EDA first-author affiliation | 13,847 | 0.39 | 0.49 | 0 | 1 |
| Senior first author, 5 or more years | 13,847 | 0.36 | 0.48 | 0 | 1 |
| Tier 1 journal | 13,847 | 0.13 | 0.33 | 0 | 1 |
| COVID period, 2020 to 2021 | 13,847 | 0.31 | 0.46 | 0 | 1 |

### 4.2 Pre-period estimates

With 2022 as the reference year, the reported group interaction is 0.003 for 2020 ($p = 0.914$) and -0.004 for 2021 ($p = 0.893$). These estimates do not show a detectable difference in group trends during the short pre-period. However, two earlier years provide little power to identify gradual divergence or other violations of the parallel-trends assumption.

### 4.3 Main estimates

Table 2 reports the main comparative pre/post models. Model 3 adds the group indicator and its interaction with Post. The Post x Non-EDA coefficient is -0.168 ($p < 0.001$). In Model 5, which includes the reported controls, the interaction is -0.176 ($p < 0.001$). With journal fixed effects in Model 6, the estimate is -0.157 ($p < 0.001$).

The Model 5 coefficient means that the post-2022 change in the non-EDA group is about 0.18 standard deviations more negative than the change in the EDA group. Expressed in the standard normal distribution, this is approximately 7 percentile points. The estimate describes a relative movement within yearly novelty distributions. It does not show that average novelty declined across the field because the outcome was standardized separately within each year.

Before 2023, the non-EDA group had higher relative semantic novelty than the EDA group. The reported Non-EDA coefficient in Model 5 is 0.201 ($p < 0.001$). The negative interaction therefore represents partial convergence between the two groups rather than evidence that non-EDA articles became uniformly less novel than EDA articles. The reported estimates are consistent with H1.

**Table 2. Comparative pre/post estimates for within-year standardized semantic novelty**

| Variable | (1) Simple | (2) Controls | (3) + Non-EDA | (4) + Senior | (5) Full | (6) + Journal FE |
|---|---|---|---|---|---|---|
| Post | 0.000 | -0.001 | 0.062* | -0.022 | 0.053 | 0.056* |
| Non-EDA | | | 0.217*** | | 0.201*** | 0.207*** |
| Senior | | | | -0.178*** | -0.145*** | -0.120*** |
| Post x Non-EDA | | | -0.168*** | | -0.176*** | -0.157*** |
| Post x Senior | | | | 0.054 | 0.019 | 0.010 |
| Tier 1 | | | | | 0.077*** | |
| Journal fixed effects | No | No | No | No | No | Yes |
| R-squared | 0.000 | 0.000 | 0.005 | 0.005 | 0.009 | 0.172 |
| N | 13,847 | 13,847 | 13,847 | 13,847 | 13,847 | 13,847 |

*Notes:* HC1 standard errors were used. The reported control set includes the log number of coauthors and the COVID-period indicator. * $p < 0.05$, ** $p < 0.01$, *** $p < 0.001$.

### 4.4 Robustness estimates

Table 3 shows that the Post x Non-EDA estimate is similar when *k* changes from 10 to 5, 15, or 20. Moving the break to 2024 produces a smaller estimate of -0.137 ($p < 0.001$). Removing the COVID years produces an estimate of -0.132 ($p = 0.015$). The placebo estimate within the pre-period is -0.021 ($p = 0.709$), which provides no evidence of a comparable shift at the false break. It does not prove that the post-2022 specification is free from pre-existing differences.

The balanced-author panel produces an estimate of -0.142 with $p = 0.104$. The balanced-author estimate is directionally similar but imprecise because the confidence indicated by the p-value includes zero. This check should be described as suggestive, not as a successful confirmation of the main result.

**Table 3. Reported robustness estimates**

| Check | Post x Non-EDA | p-value | N |
|---|---:|---:|---:|
| Main specification, k = 10 | -0.176 | < 0.001 | 13,847 |
| k = 5 | -0.178 | < 0.001 | 13,847 |
| k = 15 | -0.176 | < 0.001 | 13,847 |
| k = 20 | -0.176 | < 0.001 | 13,847 |
| Break at 2024 | -0.137 | < 0.001 | 13,847 |
| Remove 2020 and 2021 | -0.132 | 0.015 | 9,614 |
| Placebo break at 2022 | -0.021 | 0.709 | 6,415 |
| Balanced panel of repeat authors | -0.142 | 0.104 | 3,239 |

*Notes:* The specifications use within-year standardized semantic novelty and the reported controls from Model 5. The placebo uses only 2020 to 2022 observations.

Figures 1 and 2 provide descriptive views of the reported group pattern. The yearly means show that the affiliation groups moved in different directions after 2022, while the distribution plots show a larger post-period shift for the non-EDA affiliation group. These figures describe the reported data and do not identify individual LLM use or a causal effect.

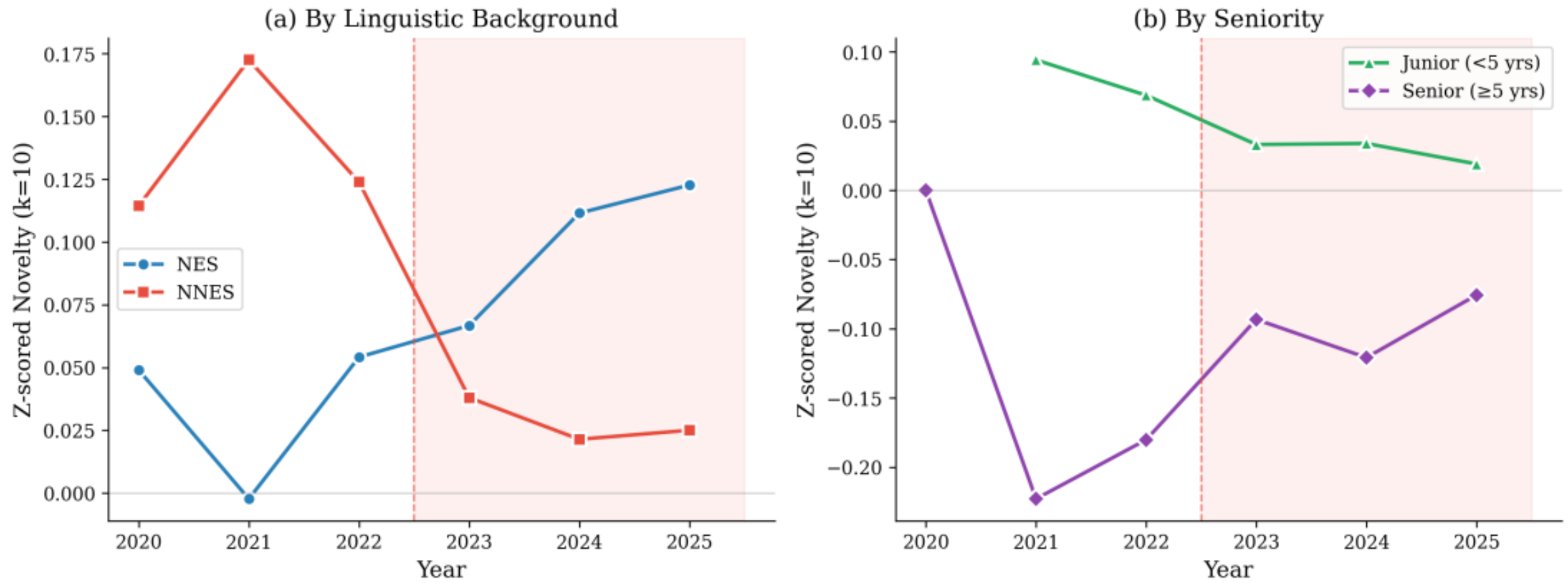

**Figure 1:** Yearly mean within-year standardized semantic novelty by affiliation context and seniority.
*Note.* In the left panel, NES and NNES denote English-dominant and non-English-dominant first-author affiliations, respectively. These groups describe institutional context, not individual language status. The shaded period begins in 2023. The plotted values are descriptive means.

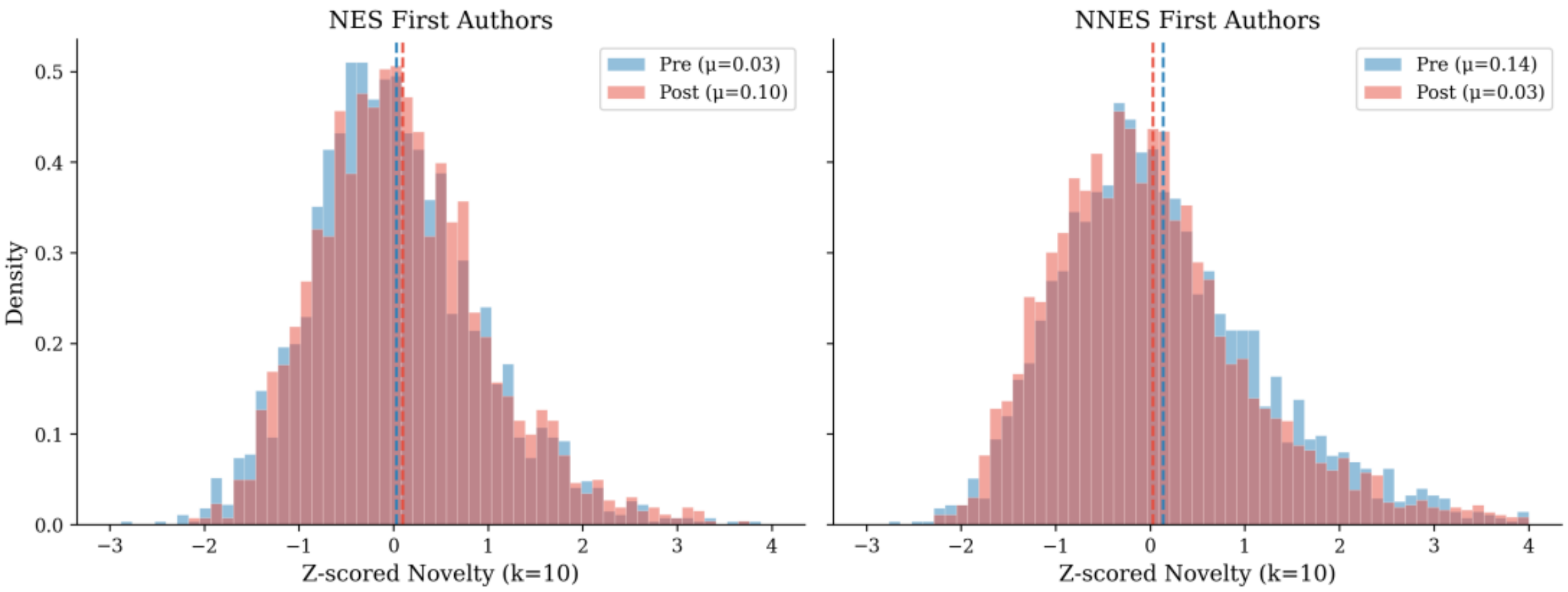

**Figure 2:** Distribution of within-year standardized semantic novelty by affiliation context and period.
*Note.* NES and NNES denote English-dominant and non-English-dominant first-author affiliations, respectively. Pre covers 2020 to 2022 and Post covers 2023 to 2025. Post is a calendar-period indicator, not a measure of LLM adoption.

# 5 Discussion

## 5.1 What the findings show

The main finding is a difference in post-2022 semantic trajectories across institutional language contexts. Articles whose first authors are affiliated with institutions in non-English-dominant countries show a larger relative decline in abstract-level semantic novelty than articles in the EDA group. The shift is meaningful in size and appears across several alternative specifications. It is also a convergence pattern. The non-EDA group begins with higher relative semantic novelty, and part of that difference becomes smaller after 2022.

The finding does not show that LLMs caused the change. It also does not show that non-EDA researchers adopted LLMs more often, depended on them more, or produced lower-quality science. The study has no direct measure of LLM use, and the affiliation variable is not an individual language measure. Several other post-2022 changes could contribute to the estimate, including shifts in topics, authors, collaboration patterns, editorial decisions, or publication timing. The careful conclusion is that a heterogeneous semantic shift occurred after ChatGPT became widely available and that the pattern is consistent with, but does not establish, a standardization process.

This interpretation is different from claiming an aggregate decline in novelty. Because the outcome is standardized within each year, the analysis cannot show whether the IS field as a whole became more or less novel. It shows how articles are positioned relative to other articles in the same year and how the relative position of the two affiliation groups changed.

## 5.2 Contribution to research on GenAI and knowledge work

The paper adds to the academic productivity literature by moving from the number of papers to a bounded feature of their content. Kwon and Yang (2025) show that publication output increased after ChatGPT and ask whether this output reflects scientific advancement. Our analysis gives a first article-level answer for IS journals, but only for semantic positioning in titles and abstracts. The observed output increase and the relative novelty shift can coexist. More papers do not automatically create more semantic diversity, but the present design cannot link the two outcomes at the individual researcher level.

The study also clarifies an important difference in how AI enters science. Liu et al. (2025) find a positive association between AI use and scientific novelty in data ending in 2019. Their measures capture papers that use AI terms, methods, or knowledge. In those papers, AI can extend analysis and support new combinations. Our study begins from a different form of exposure: the broad availability of a general-purpose language tool that can support search, framing, and writing across topics. The two findings are not necessarily inconsistent. AI can expand the technical search space when it is part of the research method, while LLM-mediated production can also make familiar representations easier to repeat.

This distinction develops the knowledge-expansion and standardization tension in GenAI-supported work. Prior IS research explains that GenAI can combine knowledge across sources and act as a knowledge co-creator (Alavi et al. 2024). The same literature also warns that reliance on accumulated knowledge can repeat old patterns and restrict new thought (Benbya et al. 2024). Our result is consistent with the standardization side of this tension, but the current variables do not test the underlying process. A stronger theory test would measure what researchers ask the tool to do, which sources the tool retrieves, how many alternatives are considered, and which model outputs enter the final paper.

### 5.3 A limited role for construal level theory

Construal level theory may offer one possible cognitive explanation for future research, but it should not carry the main claim of this study. The theory connects psychological distance in time, space, social distance, and hypotheticality with abstract or concrete mental representations (Trope and Liberman 2010). A fast and confident LLM response could reduce perceived temporal distance or hypotheticality and move attention toward immediate execution. However, effort is not one of the theory's four core distance dimensions, and the present study does not measure psychological distance or construal.

Prior work by Kim et al. (2008) shows that temporal and social distance can change the relative weight placed on desirability and feasibility. It does not directly compare novel ideas with conventional ideas. For this reason, construal level theory is best treated here as an untested explanation that could inform a later experiment. A future study could vary when and how researchers use an LLM, measure construal before and after use, and assess the novelty of both initial ideas and final written outputs.

### 5.4 Implications for research practice

The results do not support a general recommendation to avoid LLMs. These tools can improve access to information and reduce language barriers. A more useful response is to separate convenience from judgment. Researchers can use LLMs to compare multiple explanations, identify counterarguments, and locate sources, while still checking primary evidence and making the final theoretical choices themselves. Gopal et al. (2025) make a similar point by keeping source selection, validation, and intellectual responsibility with the researcher even when AI supports the work.

Editors and research institutions should also avoid treating affiliation context as proof of LLM use. The group pattern in this study cannot identify individual behavior. Policies should focus on transparent and responsible research practice across all settings rather than placing suspicion on researchers from particular countries. The observed heterogeneity is a reason to study access, use, and outcomes more carefully, not a reason to assign fault.

### 5.5 Limitations and future research

The main limitation is identification. ChatGPT's release is a common event, not a treatment assigned to one group. There is no untreated comparison group and no individual adoption measure. The interaction may reflect different exposure to LLM-related changes, but it may also reflect other group-specific changes after 2022. The results should remain associational until direct use data or a stronger source of variation becomes available.

Publication year is also a noisy time marker. Research published in 2023 or 2024 may have begun before ChatGPT. Submission and acceptance dates would create a more credible exposure window. A longer pre-period would also help assess group trends, and more post-period years would show whether the current pattern is temporary or persistent.

The outcome captures only titles and abstracts. It may reflect novelty in the research idea, novelty in how the paper is framed, or a mix of both. It does not measure novelty in theory, methods, data, or findings that is not visible in the abstract. Future work should compare several measures, including full-text semantic distance, unusual reference combinations, topic entry, and disruption. Human review of a stratified sample would also help establish whether higher embedding distance corresponds to expert judgments of novelty.

The affiliation classification is broad. It should not be interpreted as a measure of first language, writing ability, nationality, or LLM reliance. More direct measures could include educational history, self-reported language background, observed use logs collected with consent, or text-based use indicators. Even text indicators require care because editing behavior can make LLM use more or less visible across groups (Kobak et al. 2025).

The full sample may also change in composition across years. The balanced-author estimate reduces this concern but is imprecise because the sample is much smaller. A stronger analysis should examine author fixed effects, new entrants, topic composition, and journal-specific timing. The reported HC1 standard errors also do not account for possible dependence within authors or journals. Standard errors clustered at suitable levels, and sensitivity to alternative clustering choices, should be assessed when the analysis files are available.

Finally, the analysis covers one field and a selected set of journals. IS is a useful setting because it is closely connected to digital technology and includes scholars across many countries. Still, the pattern may differ in fields with other publication cycles, writing practices, or uses of LLMs.

## 6 Conclusion

This paper examines whether the semantic position of published IS research changed after ChatGPT became widely available. Across 13,847 articles, the reported results show a relative post-2022 decline in abstract-level semantic novelty for articles whose first authors are affiliated with institutions in non-English-dominant countries compared with the EDA group. The pattern is visible across several alternative specifications, although the smaller balanced-author result is not statistically precise.

The result is best read as evidence of heterogeneous change, not as a causal estimate of LLM use. It adds a careful content-based outcome to research on LLMs and academic productivity and points to a central tension in AI-supported knowledge work. GenAI can help researchers reach and combine more knowledge, but it can also make established ways of presenting that knowledge easier to repeat. Understanding which path dominates will require direct measures of tool use, stronger timing data, and broader measures of scientific novelty.